%Paper: gr-qc/9401020
%From: HARALD%CZHETH5A.BITNET@vm.cnuce.cnr.it
%Date: Wed, 19 Jan 94 14:26 N

%
\magnification=\magstep1
\input amstex
\vsize=22.5truecm
\hsize=16truecm
\hoffset=2.4truecm
\voffset=1.5truecm
\parskip=.2truecm
\font\ut=cmbx10
\font\ti=cmbx10 scaled\magstep1

\def\br{\hfill\break\noindent}
\TagsOnRight
%
% short forms of greek letters
\def\al{\alpha}
\def\be{\beta}

\def\de{\delta}

\def\la{\lambda}

\def\Om{\Omega}

% vectors
\def\vx{\vec{x}}
\def\vxp{\vec{x}\,'}
\def\vy{\vec{y}}
\def\vk{\vec{k}}
\def\vkp{\vec{k}\,'}

% derivatives
\def\vt{\!\vartriangle\!\!}

% modes and operators

% expectations, correlations, Wightman function, em--tensor

\def\rxry{\langle\rho(x)\rho(y)\rangle-\langle\rho(x)\rangle^2}

\def\dk{\langle|\,\de_\tau(k)\,|^2\rangle}

%miscellaneous

\def\bla{\big\langle}
\def\bra{\big\rangle}

\rightline{ETH-TH/93--15}
\rightline{January 1994}
\vskip2truecm
\centerline{\ti Energy Fluctuations Generated by Inflation}
\vskip2truecm
\centerline{Harald F. M\"uller and Christoph Schmid}
\vskip0.5truecm
\centerline{Institut f\"ur Theoretische Physik, ETH--H\"onggerberg,
            CH--8093 Z\"urich, Switzerland}

\vskip2.5truecm

\centerline{\ut Abstract}
\vskip.5truecm
\noindent{\sl
 The energy density correlation function
 $C(x,y)=\langle\rho(x)\rho(y)\rangle-\langle\rho\rangle^2$ and its Fourier
 transform generated by the gravitational tidal forces of the inflationary
 de Sitter expansion are derived for a massless or light ($m\ll H$) neutral
 scalar field $\Phi$ without self--interactions, minimally coupled to gravity.
 The field has no classical background component, $\langle\Phi\rangle=0$.
 Every observationally relevant mode (which has today
 $\lambda_{phys}<H_0^{-1}$) had at early times $R/k_{phys}^2\to0$, and is
 taken to be initially in the Minkowski vacuum state.
 Our computation of $C(x,y)$, which involves four field operators, is finite
 and unambiguous at each step, since we use the following two tools:
 (1) We use a normal ordered energy density operator {\rm N}$[\rho]$ and show
 that any normal ordering {\rm N} gives the same finite result.
 (2) Since $C(x,y)$ has the universal $|x-y|^{-8}$ short--distance behaviour,
 the Fourier transform can only be performed after smearing the energy density
 operator in space and time with a smearing scale $\tau$.
 The resulting energy density fluctuations are non--Gaussian, but obey a
 $\chi^2$--distribution.
 The power spectrum $\dk$ involves a smearing scale $\tau$, and we choose
 $\tau=k^{-1}$.
 For massless scalars we obtain $k^3\dk\sim H^4k^4$ on super--horizon scales
 $k < H$.
 For light scalars with masses $m\ll H$ a plateau appears on the largest
 scales $k < \sqrt{m^3/H}$. Here the result is
 $k^3\dk\sim m^6H^2 (k/H)^{4m^2/3H^2}$.
 On sub--horizon scales $k > H$ we have the universal law $k^3\dk\sim k^8$.
}

\vfill\eject

{\ti 1. Introduction and Conclusions}

Gravitational tidal forces are capable to excite a quantum field, energy
is produced, and energy fluctuations are amplified above the vacuum
fluctuations [1]. The most prominent phenomena are Hawking radiation around
a black hole and the generation of energy fluctuations in the early universe
[2,3], which serve as seeds for the formation of the observable large scale
structure [4].

The fluctuations in the energy density $\rho$ are given by the two--point
correlation function $\langle\rho\rho\rangle - \langle\rho\rangle^2$. Since
$\rho$ is bilinear in the field, the fluctuations are in general quartic in
the quantum field.

In the literature [2,3,5,6] the fluctuations in the energy density were
determined for the field which drives inflation (the inflaton field),
a minimally coupled scalar field $\chi$ with self--interactions given by
a potential $V(\chi)$. The inflationary scenario needs $\chi$ to be
dominated by a classical background component, $\langle\chi\rangle=\chi_0$,
around which the field fluctuates, $\chi=\chi_0+\de\chi$. $\chi_0$ is supposed
to be spatially homogeneous and slowly varying in time. To lowest order in
$\de\chi/\chi_0$ the fluctuations in the energy density are given by the
interference between the background $\chi_0$ and the field fluctuations
$\de\chi$, hence they are quadratic in $\de\chi$. Such a scenario gives rise
to {\it Gaussian} energy fluctuations with an approximately
{\it scale--invariant} (Harrison--Zel'dovich) spectrum and an amplitude that
is {\it much too large} unless one assumes unexpectedly small numerical values
for the parameters of the inflaton potential $V$ ({\it e.g.}
$\la\approx 10^{-12}$ for $V(\chi)\sim\la\chi^4$).

In this paper we investigate energy fluctuations of additional scalar fields
$\Phi$ that have no such background component, {\it i.e.} with
$\langle\Phi\rangle=0$. To this end we have to compute the full four field
operator quantity $\bla(\rho - \langle\rho\rangle)^2\bra$. It is the
purpose of this paper to show how this quantum field theoretical computation
can be accomplished. Therefore we keep the system under consideration as
simple as possible: We take scalar fields without self--interactions and
with minimal coupling to an external de Sitter background. We will restrict
ourselves to light scalars $\Phi$, {\it i.e.} with masses $m$
that are small compared to the Hubble constant $H$ of the background, $m\ll H$,
including the massless case, $m=0$.

The state of the quantum field is dictated by the inflationary paradigm:
(1) Every mode with comoving wave number $k$ which is of interest to physics
observable today, had a physical wavelength much smaller than the Hubble
radius, $\la_{phys}\ll H^{-1}$, if we go back early enough during the
inflationary era. At these early stages of inflation every relevant mode had
$k_{phys}^2\gg R$, {\it i.e.} the curvature $R$ was irrelevant for these
modes, they behaved as if they were in Minkowski space.
(2) According to the inflationary paradigm we want to consider the minimal
quantum fluctuations, {\it i.e.} in every relevant mode we start out with
Minkowski vacuum fluctuations. During inflation the initial fluctuations in
each mode get amplified by the gravitational tidal forces. This produces a
quantum state which is the same as the Bunch--Davies state [7] insofar as
modes with physical wavelengths today smaller than the present Hubble radius
are concerned.

The observable inhomogeneity signals in our universe give a statistical
ensemble consisting of different regions of our observable universe. However,
when the length scales in our observations approach the present Hubble radius
$H_0^{-1}$ there is no more ensemble available. Quantum field theory (for some
given model) predicts statistical averages about an ensemble of universes. If
we want to compare these predictions with observations, we must restrict the
length scales to values that are smaller than $H_0{}^{-1}$, and we postulate
the equivalence between averaging observations over different regions of our
observable universe and quantum field theoretical expectation values
(averaging over many universes). A necessary condition for this equivalence is
the removal of all modes with wavelengths $\la_{phys}>H_0{}^{-1}$ from the
quantum field theoretical expressions. This follows because such modes cannot
produce inhomogeneities on scales $\ell<H_0^{-1}$ in the correlation function
$C(\ell)$, they only can predict fluctuations for the average density of our
universe, but this is unobservable. The modes with the largest wavelengths of
interest have a physical wavelength of order $H_0{}^{-1}$ today, and at the
end of inflation they had a ratio of physical wavelength to the Hubble radius
during inflation $H^{-1}$ denoted by $H\la=e^N$. Therefore, the long
wavelength cut--off in our computation at the end of inflation will be at
$k_{min}=He^{-N}$. Assuming the present values $\Om_0=1$ and
$H_0=50\,\,km\,Mpc^{-1}\,s^{-1}$, $H=10^{16}\,\,GeV$ during inflation, and
rapid transitions from inflation to radiation and matter dominated eras we
obtain $N\approx70$. The duration of inflation $t_f-t_i$ must be at least
quite a bit longer than $NH^{-1}$, {\it i.e.} $\exp H(t_f-t_i)\gg e^N$.

Our computation is {\it finite and unambiguous at each step}, since we employ
the following two tools: (1) We use a {\it normal ordered} energy density
operator N$[\rho(x)]$, and we show that any normal ordering prescription N
gives the same finite result for the correlation function in $x$--space
$\langle\rho(x)\rho(y)\rangle - \langle\rho\rangle^2$. Beyond this normal
ordering our $x$--space calculation needs no ultraviolet regularization and
subtraction. (2) Since at short distances
$\langle\rho(x)\rho(y)\rangle - \langle\rho\rangle^2$ has the universal
$|x-y|^{-8}$ behaviour, its Fourier transform is undefined. Therefore we
consider the {\it smeared} energy density operator $\rho_\tau(x)$, where the
smearing scale in space and in time is $\tau$. The $x$--space correlation
function is regular at short distances, and its Fourier transform gives a
finite and unambiguous correlation function in $k$-space. Correlation
functions are simpler in $x$--space than in $k$--space since the former do
not need any smearing.

This has to be distinguished from the considerations in ref. [8] where it
was proposed that $\langle$N$[\rho\rho]\rangle - \langle$N$[\rho]\rangle^2$
constitutes a measure of the energy density fluctuations in a multi--particle
state in flat space. The normal ordering of the energy density bilinear in
N$[\rho^2]$ instead of considering (N$[\rho])^2$ sets the quantum fluctuations
in the Minkowski vacuum state to zero. As a general prescription this is
incorrect, since non--zero vacuum fluctuations of the energy density in the
Minkowski vacuum state are a fundamental property of quantum field theory.

The resulting energy density fluctuations are {\it non--Gaussian}, even
though the field $\Phi$ itself is Gaussian. Different $\vk$--modes
decouple in the external field approximation in FRW space--times. In each
$\vk$--mode, the field $\Phi_k$ and its canonically conjugate field $\Pi_k$
are Gaussian distributed. This is so, because the initial Gaussian character
of the distribution cannot be altered by the dynamics of free fields in
external curved space--time. Since the energy density $\rho$ is a bilinear in
the field, it cannot be a Gaussian variable. As a square of a Gaussian it
obeys a $\chi^2$--distribution. This is in contrast to the energy fluctuations
for the inflaton field, where the energy fluctuations are proportional to the
Gaussian field fluctuations.

The window size $\tau$ and the sample size $k_{min}^{-1}$ are in general
relevant physics parameters both for observations and for quantum field
theoretical predictions. The power spectrum $\dk$ necessarily contains the
smearing scale $\tau$, {\it i.e.} it is a function of two length variables,
$k^{-1}$ and $\tau$. It would be complicated to plot observations in the
two--dimensional $(\tau,k^{-1})$--plane. In the literature the two variables
are often used interchangeably, see {\it e.g.} sect. 8.1 of ref. [5]. To
present our results in a economical way we set $\tau=k^{-1}$.

For {\it massless scalars} our model contains one intrinsic scale $H$. The
power spectrum $\dk$ therefore consists of two parts: For the sub--horizon
scales, $k>H$, it varies as $k^3\dk\sim k^8$, while in the super--horizon
regime, we obtain $k^3\dk\sim H^4\,k^4$ for $k < H$. The physical implications
of this spectrum cannot be discussed, unless the inflationary phase has
terminated, and is followed by the radiation and matter dominated eras. Then
all the super--horizon scales will gradually re--enter the horizon. Recall
that a spectrum is called scale invariant (Harrison--Zel'dovich spectrum), if
$k^3\dk$ is independent of $k$ at the {\it second horizon crossing}. We shall
show in the second paper [9] that the massless scalars give rise to a scale
invariant spectrum in both radiation and matter dominated era. The amplitude
of the scale invariant part has the right order of magnitude, only if the
Hubble constant $H$ during inflation was of the order $10^{16}\,GeV$,
{\it i.e.} if the vacuum energy density during inflation was around
$(3\cdot10^{17}\,GeV)^4$.

For {\it light} scalars with masses $m^2< H^2/N$ a second scale arises,
$k_*:=m\sqrt{Nm/H}\ll H$ where $N=\log(H/k_{min})$. Therefore we must
distinguish two different regimes in the power spectrum on super--horizon
scales. On the largest scales, $k_{min}<k<k_*$, an almost flat region appears,
here we obtain $k^3\dk\sim H^4k_*^4\,(k/H)^{4m^2/3H^2}$. On scales $H<k<k_*$
the behaviour is $k^3\dk\sim H^4\,k^4\,(k/H)^{4m^2/3H^2}$.

The paper has the following structure: In the next section we derive the
two--point correlation function $C(\vx,\vxp)$ of the energy density in
$\vx$--space. In sect. 3 we show how the Fourier transform of $C(\vx,\vxp)$
can be obtained, and we derive the power spectrum $\dk$ of the energy
fluctuations.

\vskip0.5truecm
{\ti 2. Correlation Function in Position Space}

We now derive the energy--energy correlation $\rxry$ for a neutral scalar
field $\Phi$ without background component, {\it i.e.} with
$\langle\Phi\rangle=0$, without self--interactions, and with minimal coupling
to an external de Sitter space--time given by
$$
ds^2= dt^2-a(t)^2 dx_idx^i, \quad
a(t)= a(t_1)e^{H(t-t_1)}.
\tag2.1
$$
The action for the field $\Phi$ is
$$
S={1\over 2}\,\,\int\,d^4x\sqrt{-g}\,\,
 \left(\partial_\mu\Phi\partial^\mu\Phi - m^2\Phi^2\right).
\tag2.2
$$

The {\it energy density}
$\rho$ is always defined to be the one measured
by a comoving observer who in FRW coordinates has the 4--velocity
$u^\mu = (1,0,0,0)$,
$$
\rho:= T_{\mu\nu}u^\mu u^\nu = T_{00}.
\tag2.3
$$
For the neutral, minimally coupled scalar field without self--interactions
we have
$$
\aligned
\rho &= {1\over 2}\,\left[ (\partial_{\hat 0}\Phi)^2 +
           (\partial_{\Hat i}\Phi)^2 + m^2\Phi^2 \right] \\
     &= {1\over 2}\,\left[(\partial_t\Phi)^2 + {1\over a^2}(\partial_{\vx}
             \Phi)^2 + m^2\Phi^2 \right],
\endaligned
\tag2.4
$$
where $\partial_{\hat 0}$ and $\partial_{\Hat i}$ refer to
comoving local orthonormal frames (LONF).

The {\it energy--energy correlation} at equal FRW times is defined by
$$
\langle\rho(\vx)\rho(\vy)\rangle - \langle\rho(\vx)\rangle^2 =: C(\ell)
\tag2.5
$$
where $\ell$ is the measured distance between the two points $\vx$ and $\vy$.
The correlation function can only depend on $\ell$, because the quantum state
is an eigenstate of the total momentum $\vec{P}$ with eigenvalue zero.

To compute the correlation function we renormalize our energy density
operators using {\it normal ordering}, $\rho(x):= \hbox{N}[\rho(x)]$.
The key observation is that any normal ordering
prescription gives identical results for the fluctuations.
This is so because the difference between two normal orderings is a c--number,
$$
\hbox{N}_A[\rho] - \hbox{N}_B[\rho]\,\, =\,\, \hbox{c--number},
\tag2.6
$$
and this c--number drops out when computing $\rxry$. This is in agreement
with Wald's discussion in ref. [10]. $C(\ell)$ is {\it finite} by itself and
does not need any renormalization.

At this point it is convenient to introduce {\it conformal time} $\eta$, which
is always defined such that the light--cones are at $\pm 45^\circ$,
$$
ds^2 = a(\eta)^2\,(d\eta^2-dx_idx^i), \quad
d\eta= {dt\over a(t)}.
\tag2.7
$$
For an inflationary scale factor, $a(t)=a(t_1)\exp H(t-t_1)$, we
{\it fix the coordinates completely} as follows: for that point $t_1$ in the
history of the universe, for which we compute the energy--energy correlation,
we set
$$
a(t_1)=1,\,\, \eta(t_1) = -H^{-1}\quad
\Longleftrightarrow\quad
\eta = {-e^{-H(t-t_1)}\over H}, \,\,
a(\eta)= {-1\over H\eta}.
\tag2.8
$$

The {\it mode functions} in a spatially flat FRW space--time are
eigenfunctions of the comoving wave vector $\vk$ with $\vk_{phys}=\vk/a$.
Because of the translational invariance of the gravitational field in
3--space, different $k$'s decouple and $k$ is a conserved quantity. With the
ansatz for a fixed $k$
$$
\varphi_k(\eta,\vx)=: u_k(\eta)\,e^{i\vk\vx}
\tag2.9
$$
we obtain the equation of motion for the different $u_k$
$$
\left[ {d^2\over d\eta^2} - {2\over\eta}\,{d\over d\eta}
       + k^2 + {m^2\over H^2\eta^2}\right]\,\,u_k(\eta)\,\,=\,\,0.
\tag2.10
$$
Its solutions are linear combinations of Hankel functions H$^{(1,2)}_\nu$
times $\eta^{3/2}$,
$$
u_k(\eta)= const.\cdot\eta^{3/2}\cdot\hbox{H}^{(1,2)}_\nu (k\eta), \quad
     \nu = \sqrt{{9\over 4}-{m^2\over H^2}}.
\tag2.11
$$

For the {\it basis modes} $\varphi_k(\eta,\vx)$ we choose that fundamental
solution in (2.11) which has the time dependence $e^{-ik\eta}$ for
$\eta\to -\infty$, {\it i.e.} H$_\nu{}^{(2)}(k\eta)$. The
$\varphi_k(\eta,\vx)$ are those evolving modes (solutions of the dynamical
equations), which approached Minkowski single particle waves at early times
(when their physical wavelengths were much smaller than the horizon).
Therefore one could write more explicitely
$\varphi_k(\eta,\vx)=\varphi_k{}^{(in)}(\eta,\vx)$.

The {\it normalization} of the modes is
$$
(\varphi_k,\varphi_{k'}) = (2\pi)^3\,\delta^{(3)}(\vk-\vk '),
\tag2.12
$$
where the scalar product is
$$
(\varphi_k,\varphi_{k'}):=
i\int\,d^3x \,\sqrt{-\,^{(3)}\! g}\,\,
     (\varphi_k^*\,\overleftrightarrow{\partial_{\hat 0}}\,\varphi_{k'})
\tag2.13
$$
with $\sqrt{-\,^{(3)}\! g}\,=\,a^3$. The {\it normalized basis modes} are
$$
\align
\hbox{for }m=0:\quad
&\varphi_k(\eta,\vx)={H\over\sqrt{2k^3}}\, (k\eta - i)\,
 e^{i\,(\vk\vx - k\eta)},
\tag2.14a \\
\hbox{for }m\neq 0:\quad
&\varphi_k(\eta,\vx)={\sqrt{\pi}\over 2}\,H\eta^{3/2}\,
 \hbox{H}^{(2)}_\nu(k\eta)\, e^{i\vk\vx}.
\tag2.14b
\endalign
$$

The {\it mode expansion} of the field operator $\Phi(\eta,\vx)$ is
$$
\Phi(\eta,\vx)=\int\,{d^3k\over (2\pi)^3}
\left[ a_k\varphi_k(\eta,\vx) + {a_k}^\dag\varphi_k^*(\eta,\vx) \right],
\tag2.15
$$
where again one could write more explicitely $a_k=a_k^{(in)}$.
The canonical commutation relations
$
[\Phi(\vx),\Pi(\vy)]=i\delta^{(3)}(\vx - \vy),
$
where $\Pi=a^3\partial_t\Phi$,
fix the normalization factor in the commutation relations for the
annihilation and creation operators,
$$
[a_k,a_{k'}{}^\dag] = (2\pi)^3\,\delta^{(3)}(\vk-\vk ').
\tag2.16
$$

The {\it state of the quantum field} $|\Psi\rangle$ has been discussed in the
introduction. Early enough in the inflationary era each observationally
relevant mode (with $\la_0^{phys}<H_0^{-1}$ today) started out essentially in
Minkowski space ($R/k_{phys}^2\,\to\,0$), and for each of these modes the
initial state is the Minkowski vacuum. During inflation the vacuum
fluctuations get amplified by the gravitational tidal forces. In the
Heisenberg picture this state of each mode is denoted by $|0,in\rangle$, and
it is annihilated by the operators $a_k=a_k{}^{(in)}$ of eq. (2.15). We do not
make any assumptions on the quantum state for modes with
$\la_0^{phys}<H_0^{-1}$, hence
$$
a_k|\Psi\rangle = 0,\quad \hbox{for}\,\,k > k_{min}\,\,
\hbox{with}\,\,(k_{min}^{phys})_0 = H_0.
\tag2.17
$$
Insofar as observationally relevant modes are concerned the state
$|\Psi\rangle$ is equal to the Bunch--Davies state [7], denoted by
$|BD\rangle$, for which $a_k|BD\rangle=0$ for all values of $k$.

The {\it normal ordering} for the energy density operator, N$[\rho]$, can be
freely chosen for the purpose of computing correlation functions. The most
convenient choice is normal ordering in the Bunch--Davies basis, {\it i.e.}
with respect to the annihilation and creation operators in eq. (2.15) with
modes (2.14).

The $\rho\rho$ {\it correlation} is a sum of six terms, since $\rho$ is a sum
of three terms, eq. (2.4). Let us consider the contribution
$C_{mm}:=\langle\rho_m\rho_m\rangle-\langle\rho_m\rangle^2$ of the mass term
\linebreak
$\rho_m:=m^2\Phi^2/2$. It simplifies with our normal ordering adapted to the
Bunch-Davies state, since $\langle\Psi|\,\hbox{N}[\rho_m]\,|\Psi\rangle = 0$
for relevant modes. As a next step, we insert a complete set of intermediate
states, $\langle\Psi|\,\hbox{N}[\rho_m]\,|..\rangle
                                 \langle ..|\,\hbox{N}[\rho_m]\,|\Psi\rangle$,
and we note that only two--particle states can contribute, since
$\rho$ is bilinear in creation and annihilation operators. We arrive at
$$
C_{mm}(\vx,\vxp)\,\,=\,\,
 2\,\left[\,{m^2\over 2}\,
               \langle\Psi|\,\Phi(\vx)\Phi(\vxp)\,|\Psi\rangle\,\right]^2
  \,\,=\,\, {m^4\over 2}\,W(\vx,\vxp){}^2.
\tag2.18
$$
$W(\vx,\vxp):=\langle\Psi|\,\Phi(\vx)\Phi(\vxp)\,|\Psi\rangle$ is the Wightman
function for the state $|\Psi\rangle$, taken at equal times. Inserting the
mode expansion of the field operator $\Phi$ we obtain the integral
representation
$$
W(x,x')\,\,=\,\,\int\,\,{d^3k\over (2\pi)^3}\,\,\varphi_k(x)\varphi_k(x')^*
       \,\,=\,\,\int\,\,{d^3k\over (2\pi)^3}\,\, u_k(\eta)u_k(\eta')^*\cdot
                                                 e^{i\vk(\vx-\vxp)}.
\tag2.19
$$
As explained in the introduction, modes with physical wavelengths larger than
the Hubble radius today cannot contribute to observable inhomogeneity signals
today. Therefore the integral (2.19) must be cut off correspondingly at
$k_{min}=He^{-N}$ with $N\approx70$. The contributions of the other terms in
the energy density to the correlation are determined in an analogous way, and
we arrive at
$$
\aligned
C({\vx},\vxp) = \bigg[\,
   \sum_{\al,\be=0}^3\,\,&{1\over 2a(\eta)^4}\,(W(x,x')_{,\al\be'})^2\,
     +\,\sum_{\al=0}^3\,\, {m^2\over 2a(\eta)^2}\,(W(x,x')_{,\al})^2\\
    &+\,\sum_{\al=0}^3\,\, {m^2\over 2a(\eta)^2}\,(W(x,x')_{,\al'})^2\
     +\,{m^4\over 2}\,W(x,x')^2\,\bigg]\quad
                         \bigg|_{\eta=\eta'}
\endaligned
\tag2.20
$$
(we set $\eta=\eta'$ only at the end of the computation). For the
energy--energy correlations only the Wightman function $W(x,x')$ is needed,
together with the first derivatives with respect to its two arguments.
Note that the interference terms between modes with $k<k_{min}$ and with
$k>k_{min}$ in eq. (2.20) could produce inhomogeneity signals within the
Hubble radius today. But we shall not discuss such interference effects in
this paper, since they depend on the quantum state for modes with $k<k_{min}$.

We now consider a {\it massless scalar field}. Performing the integral (2.19)
for the Wightman function with $|\vk| > k_{min}$ yields
$$
\aligned
W(x,x')&\,\,=\,\,{H^2\over (2\pi)^2}
 \bigg[\,{\eta\eta'\over\vt r^2-\vt\eta^2}\,-\,
          {1\over 2}\log {k_{min}^2\over16}(\vt r^2-\vt\eta^2)\,\bigg] \\
\hbox{with}\quad
&\vt\eta:=\eta-\eta'\quad\hbox{and}\quad
 \vt r^2:=\sum_{i=1}^3\,(x_i-x_i')^2.
\endaligned
\tag2.21
$$
In eq. (2.20) only the first term survives for $m=0$, and the cut--off
$k_{min}$ drops out in the derivatives of $W$. The resulting energy--energy
correlation for $m=0$ is
$$
C(\ell)\,=\, {1\over (2\pi)^4}
 \left[\,{24\over\ell^8}+{14H^2\over\ell^6}+{3H^4\over 2\ell^4}\,\right],
\tag2.22
$$
where $\ell := a|\vx-\vxp|$ is the measured distance. At short distances,
$\ell\ll H^{-1}$, we find the universal $\ell^{-8}$ behaviour of $C(\ell)$
which follows already from dimensional considerations. At super--horizon
distance, $\ell\gg H^{-1}$, the specific choice of quantum state enters. Here
the energy correlations fall off with the fourth power of the inverse
distance, $C(\ell)\sim H^4/\ell^4$; that $C(\ell)$ contains four powers of $H$
can easily be read off eqs. (2.14) and (2.8).

We now consider {\it light scalars}, $m\ll H$. The integral (2.19) for the
Wightman function can be performed in closed form also for the massive modes
(2.14b), the Wightman function is given in terms of a hypergeometric function
(see {\it e.g.} ref. [7]). We subtract the contribution of the modes
$|\vk|<k_{min}$ and consider $k_{min}\ell\le1$, then we obtain
$$
\aligned
W(x,x')\,\,=\,\, {H^2\over (2\pi)^2}
 \bigg[\,{\eta\eta'\over\vt r^2-\vt\eta^2}\,+\,
  {3H^2\over2m^2}\bigg({4\eta\eta'\over\vt r^2-\vt\eta^2}&\bigg)^{m^2/3H^2} \\

\,&-\,{3H^2\over2m^2}\bigg({k_{min}^2\eta\eta'\over4}\bigg)^{m^2/3H^2}\,\bigg].
\endaligned
\tag2.23
$$
We insert $W$ into eq. (2.20) for the correlation function, and we neglect all
factors \linebreak
$(1+m^2/H^2)$, but keep factors $(H\ell)^{m^2/H^2}$. Using $k_{min}=He^{-N}$,
we obtain
$$
C(\ell)\,=\, {1\over (2\pi)^4}
 \bigg[\,{24\over\ell^8}+{14H^2\over\ell^6}+
          {3H^4\over 2\ell^4}\bigg({H\ell\over2}\bigg)^{-4m^2/3H^2}+
   {N^2m^4H^4\over 2}\,\al^2\,\left({H\ell\over2}\right)^{-4m^2/3H^2}\,\bigg].
\tag2.24a
$$
The dimensionless parameter $\al$ depends on the mass $m$ and the number $N$
of e--foldings
$$
\al = 1\quad\hbox{for}\quad {m^2\over H^2} < {3\over2N}
  \qquad\hbox{and}\qquad
\al = {3H^2\over2Nm^2}\quad\hbox{for}\quad
                            {m^2\over H^2} > {3\over2N}.
\tag2.24b
$$

At sub--horizon distance, $H\ell\ll 1$, we recover the universal $\ell^{-8}$
term. But on scales larger than $H^{-1}$ the behaviour depends on $m$ and $N$.
Let us first concentrate on the case $m^2/H^2 < 3/2N \ll 1$ where $\al=1$,
{\it i.e.} on masses $m < H/7$ for $N\approx 70$. Here the inverse mass
induces a second length scale $\ell_*^2:=\sqrt{3}/Nm^2$, and the
super--horizon scales divide into two domains. For $H^{-1}<\ell<\ell_*$, the
correlation falls off almost as $\ell^{-4}$, as it did in the massless case.
A correction due to the mass accelerates the decay. On scales,
$\ell_*<\ell < H^{-1} e^N$ we find an almost flat correlation
$C(\ell)\sim (H\ell)^{-4m^2/3H^2}$. If the mass is larger and lies in the
small window $H/7<m\ll H$, then $\ell_*\approx H^{-1}$ and the above
intermediate domain shrinks nearly to zero. In this case at all super--horizon
distances $1<H\ell<e^N$ the correlation function is almost flat.

The {\it mass dependence} of the correlation function for fixed scale
$\ell >H^{-1}$ shows an increase with the fourth power of the mass $m$, as
long as $m$ lies in the range \linebreak
$3/N\ell^2<m^2<3H^2/2N$. For smaller masses $C(\ell)$ is independent of $m$,
while for larger masses the correlation function decreases exponentially.

Finally we note that we recover the massless correlation function eq. (2.22)
by setting $m\to0$ in $C(\ell)$ for light scalars, eq. (2.24).
In contrast, in a de Sitter space of infinite duration in the past, where
we take into account the contribution of all modes, {\it i.e.} $N\to\infty$,
we get an additional constant term compared to eq. (2.22). This term is due to
the infrared divergence of the massless Wightman function of the Bunch--Davies
state in a perfect de Sitter space--time.

\vskip0.5truecm
{\ti 3. Correlation Function in Momentum Space}

In sect. 2 we have derived the energy--energy correlation function $C(\ell)$
at fixed time in position space. In order to make contact with the literature
[5,6] we must go to momentum space, {\it i.e.} perform the Fourier transform,
$$
C(\ell) = \int {d^3k\over(2\pi)^3}\,\,C(k)\,e^{i\vk\vec{\ell}}, \qquad
C(k)    = \int d^3\ell\,\,C(\ell)\,e^{-i\vk\vec{\ell}}.
\tag3.1
$$
Eqs. (2.22) and (2.24) tell us that $C(k)$ is ill--defined because of the
universal $\ell^{-8}$ behaviour of $C(\ell)$ for $\ell\to 0$.

To make the Fourier transform well--defined, we {\it smear out} the energy
density operator $\rho$ in space and time with window functions $F$,
$$
\aligned
\rho_{sm}(\eta,\vx)
 &:= \int d\tilde\eta\,\, F_\tau(\eta-\tilde\eta)
    \int d^3y \,\, F_\sigma(\vx-\vy)\,\,\rho(\tilde\eta,\vy)\\
\hbox{with}\quad
 &\int d\tilde\eta\,\,F_\tau(\tilde\eta) = 1\quad\hbox{and}\quad
  \int d^3y \,\,F_\sigma(\vy) = 1;
\endaligned
\tag3.2
$$
the subscripts $\tau$ and $\sigma$ indicate the mean smearing lengths. The
smearing of the energy density operator makes the product
$\rho_{sm}(x)\cdot\rho_{sm}(x)$ a well--defined operator [11]. Note that
already a smearing only in time does this job, while a smearing in space alone
is in general insufficient, since the former corresponds to an energy cut--off
in the intermediate states, while the latter constrains only the momentum
transfer. The smeared correlation function,
$$
C_{sm}(x,x') =
 \big\langle\rho_{sm}(x)\,\rho_{sm}(x')\big\rangle
\tag3.3
$$
($\langle\rho\rangle=0$ because of our normal ordering), is finite and
well--defined for all space--time points $x$ and $x'$. This mathematical
procedure can easily be motivated by recalling that any measurement process
necessary to determine the energy density has a finite resolution. The
physical processes involved in the measurement naturally define the smearing
functions $F_\tau(\tilde\eta)$ and $F_\sigma(\vy)$ in eq. (3.2). For
computational convenience we will take them as
$$
F_\tau(\tilde\eta)  :={1\over 2\pi}\,\,{\tau\over\tilde\eta^2+(\tau/2)^2},
               \qquad F_\sigma(\vy):=\de^{(3)}(\vy).
\tag3.4
$$
In the following, we will indicate the smeared quantities also by a subscript
$\tau$ in order to stress the dependence on the smearing length. The Fourier
transform in time of the smearing function $F_\tau$ is an exponential in the
energy $k=|\vk|$, $\exp(-k\tau/2)$. Therefore the Wightman function (2.19)
will contain an exponential energy cut--off $\exp(-k\tau)$. Instead of
rederiving the correlation with this cut--off, it is simpler to note that the
smeared energy density operator is
$$
\aligned
\rho_\tau(\eta,\vx)
 &=\,\,{\tau\over2\pi}\,\int d\tilde\eta\,\,
        {\rho(\eta+\tilde\eta,\vx)\over\tilde\eta^2+(\tau/2)^2} \\
 &=\,\,-{1\over2\pi i}\,\int d\tilde\eta\,\,\rho(\eta+\tilde\eta,\vx)\,
   \left[\,{1\over\tilde\eta+i\tau/2} - {1\over\tilde\eta-i\tau/2}\,\right].
\endaligned
\tag3.5
$$
To apply the calculus of residues we recall that the correlation function is
built of matrix elements $\langle0|\rho(x)|2\rangle$ and
$\langle2|\rho(x')|0\rangle$, where $|0\rangle$, $|2\rangle$ are zero-- and
two--particle states, respectively. Thus only the positive frequency parts of
$\rho(x)$ contribute, we must close the time integration contour in the lower
half of the complex plane, therefore we have
$\langle0|\rho_\tau(\eta,\vx)|2\rangle =
                         \langle0|\rho(\eta-i\tau/2,\vx)|2\rangle$.
For $\langle2|\rho(x')|0\rangle$ it is the other way round. Finally, we obtain
for the smeared correlation function
$$
\aligned
C_\tau(\eta,\vx;\eta',\vxp)
 &=\,\,\big\langle\rho(\eta-i\tau/2,\vx)\,\rho(\eta'+i\tau/2,\vxp)\big\rangle
\\
 &=\,\, C(\eta-i\tau/2,\vx\,;\,\eta'+i\tau/2,\vxp),
\endaligned
\tag3.6
$$
{\it i.e.} the smeared correlation function at equal times is identical
to the unsmeared one with an imaginary time separation $\vt\eta=-i\tau$.

Let us now apply this formula to the case of a {\it massless scalar field}.
We repeat the computation that has led to eq. (2.22), but set $\vt\eta=-i\tau$
instead of zero
$$
\aligned
C_\tau(\ell) = {1\over (2\pi)^4}\,\,
&\bigg[\,\left({24\over(\ell^2+\tau^2)^4} -
                       {128\tau^2\ell^2\over(\ell^2+\tau^2)^6}\right) \\
&+H^2\,\left({14\over(\ell^2+\tau^2)^3} - {8\tau^2\over(\ell^2+\tau^2)^4} -
                       {32\tau^2\ell^2\over(\ell^2+\tau^2)^5}\right) \\
&\qquad\qquad\qquad+H^4\,\left({3\over 2(\ell^2+\tau^2)^2} -
                       {4\tau^2\ell^2\over(\ell^2+\tau^2)^4}\right)\,\,\bigg].
\endaligned
\tag3.7
$$
On scales that are much larger than the smearing length, $\ell\gg\tau$, we
recover the unsmeared correlation function, eq. (2.22). But for $\ell\ll\tau$
the correlation function $C_\tau(\ell)$ is independent of $\ell$, the
$\ell^{-8}$, $\ell^{-6}$, and $\ell^{-4}$ poles of $C(\ell)$ have been
truncated by the smearing. The Fourier transform of $C_\tau(\ell)$ is
well--defined and yields
$$
\aligned
C_\tau(k) = {1\over 16\pi^2}\,\,
\bigg[\,{k^4\over 15\tau}+
 {H^2\over\tau^3}\,\big(1&+k\tau-{2\over3}k^2\tau^2+{1\over6}k^3\tau^3\big) \\
  &+{H^4\over\tau}\,\big(1-{1\over 2}k\tau+{1\over6}k^2\tau^2\big)\,\bigg]
\cdot\exp(-k\tau).
\endaligned
\tag3.8
$$
The cosmologically relevant term of eq. (3.8) is obtained by taking both
$\tau$ and $k^{-1}$ much larger than the Hubble radius at the end of inflation
$H^{-1}$. (Today this corresponds to $\tau$ and $k^{-1}$ that are much larger
than 1 $cm$.) Then only the $H^4$ term contributes to $C_\tau(k)$,
$$
C_\tau(k) = {1\over 16\pi^2}\,\,
     {H^4\over\tau}\,\big(1-{1\over 2}k\tau+{1\over6}k^2\tau^2\big)
                                                            \cdot\exp(-k\tau)
\qquad (k,\tau^{-1}\ll H).
\tag3.9
$$
The $1/\tau$ dependence is obvious for dimensional reasons. For
$k\ll\tau^{-1}$ the correlation becomes $k$--independent,
$C_\tau(k)\to H^4/16\pi^2\tau$, while in the other limit, $k\gg\tau^{-1}$,
it vanishes exponentially. The window size $\tau$ is a relevant parameter
both for observations and for quantum field theoretical predictions.

The {\it fluctuations of the mean energy density on a given length scale}
$\tau$ are given by
$$
\langle\rho_\tau(x)^2\rangle - \langle\rho_\tau(x)\rangle^2
   \,\,=\,\,C_\tau(\ell=0).
\tag3.10
$$
As explained above, in quantum field theory the smearing over the scale $\tau$
must be primarily done in time direction. The expectation values are
$x$--independent, because the state of the quantum field has total momentum
zero ({\it i.e.} it is translation invariant). Eq. (3.10) is analogous to the
definition of "mass fluctuations on a given scale" in chap. 8.1 of ref. [5].
The vacuum fluctuations of the mean energy density on the length scale $\tau$
in Minkowski space are obtained by putting $\ell=0$ and $\tau\ll H^{-1}$ in
eq. (3.7),
$$
\langle\rho_\tau(x)^2\rangle - \langle\rho_\tau(x)\rangle^2
   \,\,=\,\,{1\over(2\pi)^4}\,{24\over\tau^8}.
\tag3.11
$$

We introduce the operator $\de_\tau(x)$ of the energy density contrast,
$$
\de_\tau(x)\,:=\,\,
 {1\over\rho_{tot}}\,\big(\,\rho_\tau(x)-\langle\rho_\tau\rangle\,\big),
\tag3.12
$$
and consider only the contribution of massless or light scalar fields to
$\de_\tau^{tot}(x)$. The average total energy density $\rho_{tot}$ governs the
evolution of the de Sitter space, \linebreak
$\rho_{tot}=3H^2M_{Pl}^2/8\pi$. The expectation value of the square of the
density contrast is given in terms of the correlation function as
$$
\langle\,\de_\tau^2(x)\,\rangle\,\,=\,\,{C_\tau(\ell=0)\over\rho_{tot}{}^2}.
\tag3.13
$$
In the cosmologically relevant case $\tau\gg H^{-1}$ eq. (3.7) gives us the
final result for the mean density fluctuations on the length scale $\tau$
$$
\langle\,\de_\tau^2(x)\,\rangle\,\,=\,\,
 {1\over(2\pi)^4}\,{1\over\rho_{tot}{}^2}\,{3H^4\over2\tau^4}\,\,=\,\,
 {2\over3\pi^2}\,(\tau M_{Pl})^{-4}.
\tag3.14
$$

The Fourier transform in the continuum version with
$\langle\,\de_\tau(\vk)\de_\tau(\vkp)\,\rangle$ proportional to
$\de^{(3)}(\vk+\vkp)$ is less convenient than the version with a finite
volume $V$ and discrete $\vk$'s,
$$
\de_\tau(\vk)\,:=\,\,\int_V d^3x\,\,\de_\tau(x)\,{e^{-i\vk\vx}\over\sqrt{V}},
\tag3.15
$$
which allows us to look at square of $\de_\tau(\vk)$,
$$
\langle\,|\de_\tau(\vk)|^2\,\rangle\,\,=\,\,{C_\tau(k)\over\rho_{tot}{}^2}.
\tag3.16
$$
The Fourier spectrum, eq. (3.16), definitely needs smearing since otherwise
both sides of the equation are infinite. Smearing inevitably leads to the
quantity $C_\tau(k)$, which depends on two length scales. In the
cosmologically interesting case the length parameters $k^{-1},\tau$ are both
much larger than the Hubble length $H^{-1}$. As explained in the introduction
we set the two length scales $\tau$ and $k^{-1}$ to be equal, and eq. (3.9)
gives us the final result for the {\it Fourier spectrum with} $\tau=k^{-1}$
$$
\aligned
k^3\langle\,|\de_\tau(k)|^2\,\rangle\,\,=\,\,
{1\over24\pi^2e}\,{1\over\rho_{tot}{}^2}\,H^4k^4\,\,&=\,\,
{8\over27e}\,\bigg({k\over M_{Pl}}\bigg)^4, \\
 &\qquad (k=\tau^{-1}\ll H)
\endaligned
\tag3.17
$$
which is closely related to the result in eq. (3.14).

As we have seen in section 2, for {\it light scalars} with masses $m\ll H$ the
correlation function $C(\ell)$ gets a term in addition to the massless case.
It makes $C(\ell)$ almost flat on the largest scales, $\ell>\ell_*$. For this
term we could perform the Fourier transformation without any smearing of the
energy operators, {\it i.e.} there we are allowed to set $\tau=0$.
Nevertheless, in order to be consistent with the previous consideration we
again identify the two parameters as $\tau=k^{-1}$ (this alters the magnitude
just by a factor $2/e$). As in $x$--space, we obtain a plateau--like
contribution to the power spectrum on scales $k_{min}<k<k_*$ where
$k_*=\sqrt{2\al Nm^3/H}$. The result is
$$
\aligned
k^3\,\langle\,|\de_\tau(k)|^2\,\rangle\,\,=\,\,
{1\over24\pi^2e}\,{1\over\rho_{tot}{}^2}\,H^4k_*^4
                       \bigg({2k\over H}\bigg)^{4m^2/3H^2}\,\,&=\,\,
          {32N^2\al^2\over27e}\,{m^6\over H^2M_{Pl}{}^4}
                       \bigg({2k\over H}\bigg)^{4m^2/3H^2} \\
 &\qquad\qquad (k=\tau^{-1}\ll H)
\endaligned
\tag3.18
$$
where the parameter $\al$, defined in eq. (2.24b), equals unity if $m<H/7$.
On scales $k_*<k<H$ the behaviour is $k^3\dk\sim H^4k^4(k/H)^{4m^2/3H^2}$.
In the plateau region the sample size $k_{min}^{-1}$ (which determines $N$
and $k_*$) is a relevant parameter for both observations and theoretical
predictions.

The functions $C_\tau(\ell)/\rho_{tot}{}^2$ and $k^3\dk$ contain the same
information, since they are related by a Fourier transformation. In contrast
the magnitudes of these two functions at "corresponding points" $\ell=k^{-1}$
are comparable only for some special functional forms, like for
$C(\ell)\sim\ell^{-4}$, see eqs. (2.22) and (3.17). But for the almost flat
part, $C(\ell)\sim\ell^{-\epsilon}$, the magnitude of $k^3\dk$ contains an
additional factor $\epsilon\sim m^2/H^2$ in comparison with $C(\ell)$ at
corresponding points $\ell=k^{-1}$, see eqs. (2.24a) and (3.18). It follows
that the onset of the plateaus in $k^3\dk$ at $k_*$ and in $C(\ell)$ at
$\ell_*$ does not occur at corresponding points, $\ell_*\neq k_*^{-1}$.

\vskip1truecm
%Acknowledgements
We would like to thank J\"urg Fr\"ohlich, Christophe Massacand, and
Slava Mukhanov for many valuable discussions and comments.

\vfill\eject
{\ti References}
\vskip0.5truecm

\item{[1]}  N.D. Birrell and P.C.W. Davies,
                       {\sl Quantum Fields in Curved Space}, \br
                       Cambridge University, 1982; \br
            B.S. DeWitt, {\sl Phys. Rep.} {\bf 19C}(1975), 295.
\item{[2]}  K. Olive, {\sl Phys. Rep.} {\bf 190}(1990), 307; \br
            V.F. Mukhanov, H.A. Feldman, and R.H. Brandenberger, \br
                       {\sl Phys. Rep.} {\bf 215}(1992), 203.
\item{[3]}  A.H. Guth and S.-Y. Pi,
                       {\sl Phys. Rev. Lett.} {\bf 49}(1982), 1110; \br
            S.W. Hawking, {\sl Phys. Lett.} {\bf 115B}(1982), 295; \br
            A.A. Starobinsky, {\sl Phys. Lett.} {\bf 117B}(1982), 175; \br
            J.M. Bardeen, P.J. Steinhardt and M.S. Turner,
                       {\sl Phys. Rev. D} {\bf 28}(1983), 679.
\item{[4]}  G.F. Smoot, {\it et al.}, {\sl Astrophys. J. Lett.}
                        {\bf 396}(1992), L1; \br
            E.L. Wright, {\it et al.}, {\sl Astrophys. J. Lett.}
                         {\bf 396}(1992), L13.
\item{[5]}  E.W. Kolb and M.S. Turner, {\sl The Early Universe},
                       Addison--Wesley, 1990.
\item{[6]}  A.D. Linde, {\sl Particle Physics and Inflationary Cosmology},\br
                       Harwood Academic, 1990.
\item{[7]}  T.S. Bunch and P.C.W. Davies, {\sl Proc. Roy. Soc. London}
                       {\bf A360}(1978), 117.
\item{[8]}  Chung--I Kuo and L.H. Ford, {\sl Phys. Rev. D}
                       {\bf 47}(1993), 4510.
\item{[9]}  H.F. M\"uller and C. Schmid, {\sl ETH--Preprint}
                       {\it (in preparation)}.
\item{[10]} R.M. Wald, {\sl Commun. Math. Phys.} {\bf 54}(1977), 1;
                       {\sl Phys. Rev. D} {\bf 17}(1978), 1477;
                       {\sl Ann. Phys. (N.Y.)} {\bf 110}(1978), 472.
\item{[11]} See {\it e.g.} J. Glimm and A. Jaffe, {\sl Quantum Physics},
                       Springer, 1987.

\end